\documentstyle[12pt]{article}
\newcommand{\definition}{\stackrel{\rm d}{\equiv}}
\newcommand{\vecvar}[1]{\mbox{\boldmath$#1$}}
\begin{document}
\begin{titlepage}
\begin{flushright}
TOYAMA-84 \makebox[2mm]{} \\
\makebox[2mm]{}
\vspace{3cm}\\
\end{flushright}
\begin{center}
{\bf\Huge Gauge Theory of Massive Tensor Field}
\vspace{4cm}\\
{\Large Shinji HAMAMOTO}
\vspace{1cm}\\
{\it Department of Physics, Toyama University \\
Toyama 930, JAPAN}
\vspace{3cm}\\
\end{center}
\begin{abstract}
In order to construct a massive tensor theory with a smooth massless limit,
we apply the Batalin-Fradkin algorithm to the ordinary massive tensor
theory.
By introducing an auxiliary vector field all second-class constraints
are converted into first-class ones.
We find a gauge-fixing condition which produces a massive tensor theory
of desirable property.
\end{abstract}
\end{titlepage}

\section{Introduction}

The purpose of the present paper is to construct a massive tensor field
theory with a smooth massless limit.
For a massive vector field it has been known the Stueckelberg
formalism has a smooth massless limit.
For a massive tensor field Fronsdal and Heidenreich (FH) succeeded in
constructing such a theory.${}^{1)}$
Based on the Lagrangian formalism, they removed massless singularities
from all two-point functions
by introducing two kinds of auxiliary fields with spin-1 and 0.
In the present paper we take a different way from
FH for the same purpose.
Our method based on the Hamiltonian formalism is more systematic than
FH's.

We apply the Batalin-Fradkin (BF) algorithm${}^{2),3)}$ to the ordinary
massive tensor theory.
By introducing only a spin-1 auxiliary field (BF field),
contrary to the case of FH,
we convert the original second-class constrained system into a
first-class one.
To the resulting gauge-invariant system two kinds of gauge-fixings
are imposed.
One is unitary gauge-fixing that eliminates the BF field from the system
to recover the
original massive tensor theory.
The other is massless-regular gauge-fixing that leads to a theory with
a smooth massless limit.
Based on the Hamiltonian formalism, however, our formulation is
non-covariant from the beginning.
We have not succeeded in getting final results in covariant form yet.
In this regard we are below FH,
while the number of auxiliary fields introduced here is smaller than in FH.
To get some covariant expressions is to be left for future studies.

In \S 2, following Igarashi,${}^{4)}$
we review the case of a massive vector field.
This is to see how the Stueckelberg formalism follows from applying BF
algorithm to a massive vector theory.
Section 3 treats the usual canonical formalism of a massless tensor field.
The canonical formalism of a massive tensor field is presented in \S 4.
In \S 5 BF algorithm is applied to this system.
By introducing an auxiliary vector field, the original second-class
constrained system is converted into a first-class one.
In \S 6 two kinds of gauge-fixings, unitary one and massless-regular one,
are investigated.
It is shown that the former recovers the original system,
while the latter gives a system with a smooth massless limit.
Section 7 is devoted to summary and discussion.

\section{Massive vector field}

\renewcommand{\thefootnote}{\fnsymbol{footnote}}
A massive vector field is described by the Lagrangian
\footnote[2]
{
In the present paper Greek indices run $0-3$, while Latin indices $1-3$.
The metric is
$\eta^{\mu\nu} \definition ( -1, +1, +1, +1 )$.
}
\begin{equation}
L[A] =
- \frac{1}{4}F_{\mu\nu}F^{\mu\nu} - \frac{m^{2}}{2}A_{\mu}A^{\mu} ,
\end{equation}
where
\begin{equation}
F_{\mu\nu} \definition \partial_{\mu}A_{\nu} - \partial_{\nu}A_{\mu} .
\end{equation}
This system has a primary constraint
{
\setcounter{enumi}{\value{equation}}
\addtocounter{enumi}{1}
\setcounter{equation}{0}
\renewcommand{\theequation}{\theenumi\alph{equation}}
\begin{equation}
\varphi_{1} \definition \pi^{0} \approx 0
\end{equation}
and a secondary constraint
\begin{equation}
\varphi_{2} \definition \partial_{i}\pi^{i} + m^{2}A_{0} \approx 0 ,
\end{equation}
where $\pi^{\mu}$ are momenta conjugate to $A_{\mu}$.
The Poisson bracket between the constraints
\setcounter{equation}{\value{enumi}}
}
\begin{equation}
[ \varphi_{1}(\vecvar{x}), \varphi_{2}(\vecvar{x}') ]
= - m^{2}\delta^{3}(\vecvar{x} - \vecvar{x}')
\end{equation}
shows that the constraints are of the second class when the mass $m$ is
finite.
The Hamiltonian is calculated to be
\begin{equation}
H =
\frac{1}{2}\pi^{i2} + \frac{1}{4}F_{ij}^{\makebox[2mm]{}2}
+ \frac{m^{2}}{2}(A_{0}^{\makebox[1mm]{}2} + A_{i}^{\makebox[1mm]{}2})
+ \partial^{i}A_{i}\varphi_{1} - A_{0}\varphi_{2} .
\end{equation}
The time developments of the constraints are
\begin{equation}
\begin{array}{rcccl}
\dot{\varphi}_{1} & = & [\varphi_{1}, H] & = & \varphi_{2} , \\
\dot{\varphi}_{2} & = & [\varphi_{2}, H] & = &
\partial_{i}^{\makebox[1mm]{}2}\varphi_{1} .
\end{array}
\end{equation}

In order to convert the above second-class constrained system into
a first-class one, we introduce a BF field $\theta$ and its conjugate
momentum $\omega$, modifying the constraints and Hamiltonian as follows:
\begin{equation}
\begin{array}{rclcl}
\tilde{\varphi}_{1} & \definition & \tilde{\pi}^{0} & \approx & 0 , \\
\tilde{\varphi}_{2} & \definition &
\partial_{i}\pi^{i} + m^{2}\tilde{A}_{0} & \approx & 0 ;
\end{array}
\end{equation}
\begin{equation}
\tilde{H} \definition
\frac{1}{2}\pi^{i2}
+ \frac{1}{4}\tilde{F}_{ij}^{\makebox[2mm]{}2}
+ \frac{m^{2}}{2}
   (\tilde{A}_{0}^{\makebox[1mm]{}2} + \tilde{A}_{i}^{\makebox[1mm]{}2})
+ \partial^{i}\tilde{A}_{i}\tilde{\varphi}_{1}
- \tilde{A}_{0}\tilde{\varphi}_{2} ,
\end{equation}
where
\begin{equation}
\begin{array}{rcl}
\tilde{A}_{0} & \definition & A_{0} + \theta , \\
\tilde{A}_{i} & \definition & \displaystyle
A_{i} - \frac{1}{m^{2}}\partial_{i}\omega , \\
\tilde{\pi}^{0} & \definition & \pi^{0} - \omega .
\end{array}
\end{equation}
The resulting system is in fact of the first class:
\begin{eqnarray}
\left[ \tilde{\varphi}_{1}, \tilde{\varphi}_{2} \right] & = & 0 , \\
\left[ \tilde{\varphi}_{\alpha}, \tilde{H} \right] & = & 0 .
\makebox[1cm]{} ( \alpha = 1, 2 )
\end{eqnarray}

Now we come to the gauge-fixings.
The first choice is the ^^ unitary gauge':
\begin{equation}
\begin{array}{rclcl}
\chi^{1} & \definition & \theta & \approx & 0 , \\
\chi^{2} & \definition & \displaystyle \frac{1}{m^{2}}\omega & \approx
& 0 .
\end{array}
\end{equation}
For this choice we get
\begin{equation}
\left[ \chi^{\alpha}(\vecvar{x}), \tilde{\varphi}_{\beta}(\vecvar{x}')
\right]
= - \delta^{\alpha}_{\beta}\delta^{3}(\vecvar{x} - \vecvar{x}') .
\end{equation}
The path integral expression is given by
\begin{equation}
Z =
\int{\cal D}\pi^{\mu}{\cal D}A_{\mu}{\cal D}\omega{\cal D}\theta
\delta(\chi^{\alpha})\delta(\tilde{\varphi}_{\alpha})
\exp i\int d^{4}x
\left[ \pi^{\mu}\dot{A}_{\mu} + \omega\dot{\theta} - \tilde{H} \right] .
\end{equation}
The variables $\pi^{0}, \theta$ and $\omega$ are integrated out,
and the factor $\delta(\partial_{i}\pi^{i} + m^{2}A_{0})$ is exponentiated
to give
\begin{eqnarray}
Z & = &
\int{\cal D}\pi^{i}{\cal D}A_{i}{\cal D}A_{0}{\cal D}\lambda
\nonumber \\
& & \times
\exp i\int d^{4}x
\left[ \pi^{i}\dot{A}_{i} - \frac{1}{2}\pi^{i2}
- \frac{1}{4}F_{ij}^{\makebox[2mm]{}2}
- \frac{m^{2}}{2}( A_{i}^{\makebox[1mm]{}2} + A_{0}^{\makebox[1mm]{}2} )
+ \lambda ( \partial_{i}\pi^{i} + m^{2}A_{0} ) \right] . \nonumber \\
& &
\end{eqnarray}
Integrating over the variables $\pi^{i}$ and $A_{0}$ and overwriting
$A_{0}$ on $\lambda$, we obtain
\begin{equation}
Z = \int{\cal D}A_{\mu}\exp i\int d^{4}x
\left[ -\frac{1}{4}F_{\mu\nu}F^{\mu\nu}
- \frac{m^{2}}{2}A_{\mu}A^{\mu} \right] .
\end{equation}
This is nothing but the original system. That means the original system (1)
can be regarded as a gauge-fixed version of the extended gauge system
(7) and (8)

Another gauge-fixing we consider is
\begin{equation}
\begin{array}{rclcl}
\chi^{1} & \definition & A_{0} & \approx & 0 , \\
\chi^{2} & \definition & \partial^{i}A_{i} & \approx & 0 .
\end{array}
\end{equation}
In this case we have
\begin{equation}
\left[ \chi^{\alpha}(\vecvar{x}), \tilde{\varphi}_{\beta}(\vecvar{x}')
\right] = \left(
\begin{array}{cc}
\delta^{3}(\vecvar{x} - \vecvar{x}') & \\
& -\partial_{i}\partial^{i}\delta^{3}(\vecvar{x} - \vecvar{x}')
\end{array}
\right) .
\end{equation}
The path integral expression is
\begin{equation}
Z =
\int{\cal D}\pi^{\mu}{\cal D}A_{\mu}{\cal D}\omega{\cal D}\theta
\delta(\chi^{\alpha})\delta(\tilde{\varphi}_{\alpha})
\prod_{t}{\rm Det}M
\exp i\int d^{4}x \left[ \pi^{\mu}\dot{A}_{\mu} + \omega\dot{\theta}
- \tilde{H} \right] ,
\end{equation}
where
\begin{equation}
M \definition
\partial_{i}\partial^{i}\delta^{3}( \vecvar{x} - \vecvar{x}' ) .
\end{equation}
The integrations over $\pi^{0}$ and $A_{0}$
and the exponentiation of the factor
$\delta ( \partial_{i}\pi^{i} + m^{2}\theta )$ are easily carried out:
\begin{eqnarray}
Z & = &
\int{\cal D}\pi^{i}{\cal D}A_{i}{\cal D}\omega{\cal D}\theta
{\cal D}\lambda\delta(\partial^{i}A_{i})\prod_{t}{\rm Det}M \nonumber \\
& & \times
\exp i\int d^{4}x\left[ \pi^{i}\dot{A}_{i} + \omega\dot{\theta}
- \frac{1}{2}\pi^{i2}
- \frac{1}{4}F_{ij}^{\makebox[2mm]{}2}
- \frac{m^{2}}{2}(A_{i}-\frac{1}{m^{2}}\partial_{i}\omega)^{2}
\right. \nonumber \\
& & \makebox[60mm]{} \left.
- \frac{m^{2}}{2}\theta^{2}
+ \lambda(\partial_{i}\pi^{i}+m^{2}\theta)\right] .
\end{eqnarray}
Integrating out $\pi^{i}$ and $\theta$ and writing $A_{0}$ and $\omega$
over $\lambda$ and $\frac{1}{m^{2}}\omega$ respectively, we obtain
\begin{equation}
Z =
\int{\cal D}A_{\mu}{\cal D}\omega\delta(\partial^{i}A_{i})
\prod_{t}{\rm Det}M\exp i\int d^{4}x L[A, \omega] ,
\end{equation}
where
\begin{equation}
L[A, \omega] \definition
- \frac{1}{4}F_{\mu\nu}F^{\mu\nu}
- \frac{m^{2}}{2}( A_{\mu} - \partial_{\mu}\omega )^{2} .
\end{equation}
Note that the Lagrangian $L[A, \omega]$ is invariant under the gauge
transformation with an arbitrary function $\varepsilon (x)$:
\begin{equation}
\begin{array}{rcl}
\delta A_{\mu} & = & \partial_{\mu}\varepsilon , \\
\delta\omega & = & \varepsilon .
\end{array}
\end{equation}
Considering this fact, the factor
\begin{equation}
\delta(\partial^{i}A_{i})\prod_{t}{\rm Det}M
\end{equation}
can be replaced by the covariant expression
\begin{equation}
\delta(\partial^{\mu}A_{\mu}-f){\rm Det}N ,
\end{equation}
where $N$ is defined by
\begin{equation}
N \definition \partial_{\mu}\partial^{\mu}\delta^{4}(x-x') ,
\end{equation}
and $f$ is an arbitrary function of $x$.
By using the Nakanishi-Lautrup (NL) field $B$ and the Faddeev-Popov (FP)
ghosts $( c, \bar{c} )$,
we can exponentiate the factor (26) to get the final
form of the path integral
\begin{eqnarray}
Z & = &
\int{\cal D}A_{\mu}{\cal D}\omega{\cal D}B{\cal D}c{\cal D}\bar{c}
\nonumber \\
& & \makebox[15mm]{} \times
\exp i\int d^{4}x\left[ L[A, \omega]
+ B(\partial^{\mu}A_{\mu} + \frac{\alpha}{2}B)
+ i\bar{c}\partial_{\mu}\partial^{\mu}c \right] ,
\end{eqnarray}
where $\alpha$ is an arbitrary constant, gauge parameter.
The transformation to the Stueckelberg formalism has been completed.
The BF field $\omega$ is now identified with the Stueckelberg field.
The expression (28) has a smooth massless limit: when the mass $m$
tends to zero, the field $\omega$ becomes redundant and the expression (28)
goes to the usual one for the Abelian gauge theory
\begin{equation}
Z = \int{\cal D}A_{\mu}{\cal D}B{\cal D}c{\cal D}\bar{c}
\exp i\int d^{4}x\left[
- \frac{1}{4}F_{\mu\nu}F^{\mu\nu}
+ B(\partial^{\mu}A_{\mu} + \frac{\alpha}{2}B)
+ i\bar{c}\partial_{\mu}\partial^{\mu}c \right] .
\end{equation}
In this respect we call the gauge choice (17) ^^ massless-regular
gauge'.

\section{Massless tensor field}

The Lagrangian is
\begin{equation}
L[h, m=0] = - \frac{1}{2}(
\partial_{\lambda}h_{\mu\nu}\partial^{\lambda}h^{\mu\nu} -
\partial_{\lambda}h^{\mu}_{\makebox[1mm]{}\mu}
\partial^{\lambda}h^{\nu}_{\makebox[1mm]{}\nu} ) +
\partial_{\lambda}h_{\mu\nu}\partial^{\nu}h^{\mu\lambda} -
\partial_{\mu}h^{\mu\nu}\partial_{\nu}h^{\lambda}_{\makebox[1mm]{}\lambda} .
\end{equation}
This system has two primary constraints
{
\setcounter{enumi}{\value{equation}}
\addtocounter{enumi}{1}
\setcounter{equation}{0}
\renewcommand{\theequation}{\theenumi\alph{equation}}
\begin{equation}
\begin{array}{rclcl}
\varphi^{0} & \definition & \pi^{0} + \partial^{m}h_{m} & \approx & 0 , \\
\varphi^{m} & \definition &
\pi^{m} - \partial^{m}h_{0} - \partial^{m}h^{n}_{\makebox[1mm]{}n}
& \approx & 0 ,
\end{array}
\end{equation}
and two secondary constraints
\begin{equation}
\begin{array}{rclcl}
\varphi^{0}_{1} & \definition &
\partial_{m}\partial^{m}h^{n}_{\makebox[1mm]{}n}
- \partial^{m}\partial^{n}h_{mn} & \approx & 0 , \\
\varphi^{m}_{1} & \definition &
\partial_{n}\pi^{mn} + \partial^{m}\partial^{n}h_{n} & \approx & 0 ,
\end{array}
\end{equation}
where $\pi^{0}, \pi^{m}$ and $\pi^{mn}$ are momenta conjugate to
$h_{0} \definition h_{00}, h_{m} \definition h_{0m}$ and $h_{mn}$.
The Poisson brackets between these constraints
$\varphi^{A} \definition
( \varphi^{0}, \varphi^{m}, \varphi^{0}_{1}, \varphi^{m}_{1} )$
are
\setcounter{equation}{\value{enumi}}
}
\begin{equation}
\left[ \varphi^{A}(\vecvar{x}), \varphi^{B}(\vecvar{x}') \right] = 0 .
\end{equation}
The Hamiltonian is
\begin{equation}
H = H_{0}(m=0) + \lambda_{A}\varphi^{A} ,
\end{equation}
where $H_{0}(m=0)$ is defined by
\begin{eqnarray}
H_{0}(m=0) & \definition &
\frac{1}{2}\pi^{mn}\pi_{mn} -
\frac{1}{4}\pi^{m}_{\makebox[1mm]{}m}\pi^{n}_{\makebox[1mm]{}n} +
2\pi^{mn}\partial_{m}h_{n} -
\frac{1}{2}\pi^{m}_{\makebox[1mm]{}m}\partial^{n}h_{n}
\nonumber \\
& & + 2\partial_{m}h_{n}\partial^{n}h^{m} -
\frac{3}{4}\partial^{m}h_{m}\partial^{n}h_{n} +
\partial_{m}h_{0}\partial^{m}h^{n}_{\makebox[1mm]{}n} -
\partial_{m}h_{0}\partial_{n}h^{mn}
\nonumber \\
& & + \frac{1}{2}\left( \partial_{l}h_{mn}\partial^{l}h^{mn} -
\partial_{l}h^{m}_{\makebox[1mm]{}m}
\partial^{l}h^{n}_{\makebox[1mm]{}n}\right) -
\partial_{l}h_{mn}\partial^{n}h^{ml} +
\partial_{m}h^{mn}\partial_{n}h^{l}_{\makebox[1mm]{}l} , \nonumber \\
& &
\end{eqnarray}
and $\lambda_{A}$ are arbitrary coefficients.
The time developments of the constraints are
\begin{equation}
\begin{array}{rclcl}
\dot{\varphi}^{0} & = & [ \varphi^{0}, H ] & = & \varphi^{0}_{1} , \\
\dot{\varphi}^{m} & = & [ \varphi^{m}, H ] & = & 2\varphi^{m}_{1} , \\
\dot{\varphi}^{0}_{1} & = & [ \varphi^{0}_{1}, H ] & = &
- \partial_{m}\varphi^{m}_{1} , \\
\dot{\varphi}^{m}_{1} & = & [ \varphi^{m}_{1}, H ] & = & 0 .
\end{array}
\end{equation}
Equations (32) and (35) show that this system is a first-class constrained
system.

In order to make the constraints second-class,
we impose the gauge-fixing conditions
$\chi_{A} \definition
( \chi_{0}, \chi_{m}, \chi_{10}, \chi_{1m} )$:
\begin{equation}
\begin{array}{rclcl}
\chi_{0} & \definition & h_{0} & \approx & 0 , \\
\chi_{m} & \definition & h_{m} & \approx & 0 , \\
\chi_{10} & \definition & \pi^{m}_{\makebox[1mm]{}m} & \approx & 0 , \\
\chi_{1m} & \definition & \displaystyle
\partial^{n}h_{mn} - \frac{1}{2}\partial_{m}h^{n}_{\makebox[1mm]{}n} &
\approx & 0 .
\end{array}
\end{equation}
The path integral expression is
\begin{eqnarray}
Z & = &
\int{\cal D}\pi^{0}{\cal D}\pi^{m}{\cal D}\pi^{mn}
{\cal D}h_{0}{\cal D}h_{m}{\cal D}h_{mn}\delta(\varphi^{A})\delta(\chi_{A})
\prod_{t}{\rm Det}M \nonumber \\
& & \makebox[15mm]{}\times\exp i\int d^{4}x \left[
\pi^{0}\dot{h}_{0} + \pi^{m}\dot{h}_{m} + \pi^{mn}\dot{h}_{mn} - H \right] ,
\end{eqnarray}
where
\begin{equation}
M \definition \delta_{\alpha}^{\beta}\partial_{m}\partial^{m}
\delta^{3}( \vecvar{x} - \vecvar{x}' ) .
\makebox[1cm]{}( \alpha, \beta = 0-3 )
\end{equation}
The integrations over $h_{0}, h_{m}, \pi^{0}$ and $\pi^{m}$ are easily
carried out.
The factors
$\delta( \partial_{m}\partial^{m}h^{n}_{\makebox[1mm]{}n} -
\partial^{m}\partial^{n}h_{mn} )$ and
$\delta( \partial_{n}\pi^{mn} )$ are exponentiated by
\begin{equation}
\delta( \partial_{m}\partial^{m}h^{n}_{\makebox[1mm]{}n} -
\partial^{m}\partial^{n}h_{mn} ) =
\int{\cal D}h_{0}\exp i\int d^{4}x
\left[ - \partial_{m}h_{0}\partial^{m}h^{n}_{\makebox[1mm]{}n} +
\partial_{m}h_{0}\partial_{n}h^{mn}\right] ,
\end{equation}
\begin{equation}
\delta( \partial_{n}\pi^{mn} ) =
\int{\cal D}h_{m}\exp i\int d^{4}x
\left[ - \pi^{mn}( \partial_{m}h_{n} + \partial_{n}h_{m} )\right] .
\end{equation}
We further integrate over $\pi^{mn}$ to get
\begin{eqnarray}
Z & = &
\int{\cal D}h_{0}{\cal D}h_{m}{\cal D}h_{mn}
\delta ( \partial^{n}h_{mn} -
\frac{1}{2}\partial_{m}h^{n}_{\makebox[1mm]{}n} )
\prod_{t}{\rm Det}M \nonumber \\
& & \times\exp i\int d^{4}x
\left[ L[h, m=0] + \frac{4}{3} ( \partial^{m}h_{m} -
\frac{1}{2}\dot{h}^{m}_{\makebox[1mm]{}m} )^{2} \right] .
\end{eqnarray}
Considering the fact that the Lagrangian $L[h, m=0]$ is invariant
(up to total divergence) under the gauge transformation with four
arbitrary functions $\varepsilon_{\mu}(x)$
\begin{equation}
\delta h_{\mu\nu} =
\partial_{\mu}\varepsilon_{\nu} + \partial_{\nu}\varepsilon_{\mu} ,
\end{equation}
we can give various expressions to the generating functional $Z$.
For example, for ^^ Coulomb-like gauge' we have
\begin{eqnarray}
Z & = &
\int{\cal D}h_{0}{\cal D}h_{m}{\cal D}h_{mn}
\delta(\partial^{m}h_{m} - \frac{1}{2}\dot{h}^{m}_{\makebox[1mm]{}m} -
f_{0})
\delta(\partial^{n}h_{mn} -
\frac{1}{2}\partial_{m}h^{n}_{\makebox[1mm]{}n} - f_{m})
\prod_{t}{\rm Det}M \nonumber \\
& & \makebox[1cm]{}
\times\exp i\int d^{4}x L[h, m=0] \\
& = &
\int{\cal D}h_{0}{\cal D}h_{m}{\cal D}h_{mn}
\delta(\partial^{n}h_{mn} -
\frac{1}{2}\partial_{m}h^{n}_{\makebox[1mm]{}n})
\prod_{t}{\rm Det}M \nonumber \\
& & \makebox[1cm]{}
\times\exp i\int d^{4}x \left[ L[h, m=0] +
\frac{1}{2\alpha}(\partial^{m}h_{m} -
\frac{1}{2}\dot{h}^{m}_{\makebox[1mm]{}m})^{2}\right] ,
\end{eqnarray}
where $f_{\mu} ( \mu = 0-3 )$ are arbitrary functions of $x$,
and $\alpha$ is an arbitrary constant, gauge parameter.
The expression (41) is a special case of (44).
The covariant expressions are also obtained as follows:
\begin{eqnarray}
Z & = &
\int{\cal D}h_{\mu\nu}
\delta(\partial^{\nu}h_{\mu\nu} -
\frac{1}{2}\partial_{\mu}h^{\nu}_{\makebox[1mm]{}\nu} - f_{\mu})
{\rm Det}N\exp i\int d^{4}x L[h, m=0] \\
& = &
\int{\cal D}h_{\mu\nu}{\cal D}B_{\mu}{\cal D}c_{\mu}{\cal D}\bar{c}^{\mu}
\nonumber \\
& &
\times\exp i\int d^{4}x\left[ L[h, m=0] +
B^{\mu}(\partial^{\nu}h_{\mu\nu} -
\frac{1}{2}\partial_{\mu}h^{\nu}_{\makebox[1mm]{}\nu}) +
\frac{\alpha}{2}B^{\mu}B_{\mu} +
i\bar{c}^{\mu}\partial_{\nu}\partial^{\nu}c_{\mu}\right] , \nonumber \\
& &
\end{eqnarray}
where $N$ is defined by
\begin{equation}
N \definition \delta_{\alpha}^{\beta}\partial_{\mu}\partial^{\mu}
\delta(x-x') , \makebox[2cm]{}( \alpha , \beta = 0-3 )
\end{equation}
and NL field $B_{\mu}$ and
FP ghosts $( c_{\mu}, \bar{c}^{\mu} )$ have been introduced.

\section{Massive tensor field}

A massive tensor field is described by the Lagrangian
\begin{equation}
L[h] = L[h, m=0] - \frac{m^{2}}{2}( h_{\mu\nu}h^{\mu\nu} -
h^{\mu}_{\makebox[1mm]{}\mu}h^{\nu}_{\makebox[1mm]{}\nu} ) .
\end{equation}
This system has the same primary constraints as (31a)
{
\setcounter{enumi}{\value{equation}}
\addtocounter{enumi}{1}
\setcounter{equation}{0}
\renewcommand{\theequation}{\theenumi\alph{equation}}
\begin{equation}
\begin{array}{rclcl}
\varphi^{0} & \definition &
\pi^{0} + \partial^{m}h_{m} & \approx & 0 , \\
\varphi^{m} & \definition &
\pi^{m} - \partial^{m}h_{0} - \partial^{m}h^{n}_{\makebox[1mm]{}n}
& \approx & 0 ,
\end{array}
\end{equation}
but the different secondary constraints from (31b)
\begin{equation}
\begin{array}{rclcl}
\varphi_{1}^{0} & \definition &
\partial_{m}\partial^{m}h^{n}_{\makebox[1mm]{}n} -
\partial^{m}\partial^{n}h_{mn} -
m^{2}h^{m}_{\makebox[1mm]{}m} & \approx & 0 , \\
\varphi_{1}^{m} & \definition &
\partial_{n}\pi^{mn} + \partial^{m}\partial^{n}h_{n} + m^{2}h^{m}
& \approx & 0 , \\
\varphi_{2}^{0} & \definition &
\pi^{m}_{\makebox[1mm]{}m} + \partial^{m}h_{m} & \approx & 0 .
\end{array}
\end{equation}
The Poisson brackets between these constraints are calculated as
\setcounter{equation}{\value{enumi}}
}
\begin{equation}
\begin{array}{rcl}
\left[ \varphi^{m}(\vecvar{x}), \varphi_{1}^{n}(\vecvar{x}')\right]
& = & - m^{2}\eta^{mn}\delta^{3}(\vecvar{x}-\vecvar{x}') , \vspace{2mm}\\
\left[ \varphi^{m}(\vecvar{x}), \varphi_{2}^{0}(\vecvar{x}')\right]
& = & - 2\partial^{m}\delta(\vecvar{x}-\vecvar{x}') , \vspace{2mm}\\
\left[ \varphi_{1}^{0}(\vecvar{x}), \varphi_{1}^{m}(\vecvar{x}')\right]
& = & m^{2}\partial^{m}\delta(\vecvar{x}-\vecvar{x}') , \vspace{2mm}\\
\left[ \varphi_{1}^{0}(\vecvar{x}), \varphi_{2}^{0}(\vecvar{x}')\right]
& = &
( 2\partial_{m}\partial^{m} - 3m^{2} )\delta(\vecvar{x}-\vecvar{x}') ,
\vspace{2mm}\\
\mbox{The others} & = & 0 .
\end{array}
\end{equation}
The Hamiltonian is
\begin{equation}
H = H_{0} + \lambda_{0}\varphi^{0} + \partial^{n}h_{mn}\varphi^{m} +
( h_{0} - h^{m}_{\makebox[1mm]{}m} )\varphi_{1}^{0} ,
\end{equation}
where $H_{0}$ is defined by
\begin{equation}
H_{0} = H_{0}(m=0) + \frac{m^{2}}{2}(
h_{mn}h^{mn} - h^{m}_{\makebox[1mm]{}m}h^{n}_{\makebox[1mm]{}n} -
2h_{m}h^{m} + 2h_{0}h^{m}_{\makebox[1mm]{}m} ) ,
\end{equation}
and $\lambda_{0}$ is an arbitrary coefficient.
The Poisson brackets between the constraints and Hamiltonian are
\begin{equation}
\begin{array}{rcl}
\left[ \varphi^{0}, H \right] & = & 0 , \vspace{2mm} \\
\left[ \varphi^{m}, H \right] & = & 2\varphi_{1}^{m} , \vspace{2mm} \\
\left[ \varphi_{1}^{0}, H \right] & = & \displaystyle
- \partial_{m}\varphi_{1}^{m} + \frac{m^{2}}{2}\varphi_{2}^{0} ,
\vspace{2mm}\\
\left[ \varphi_{1}^{m}, H \right] & = & \displaystyle
\partial^{m}\varphi_{1}^{0} +
\frac{1}{2}\partial_{n}\partial^{n}\varphi^{m} +
\frac{1}{2}\partial^{m}\partial_{n}\varphi^{n} , \vspace{2mm} \\
\left[ \varphi_{2}^{0}, H \right] & = &
\partial_{m}\varphi^{m} + 4\varphi_{1}^{0} .
\end{array}
\end{equation}
Equations (50) and (53) show that $\varphi^{0}$ is a first-class
constraint and
the other constraints are of the second class.

In order to make $\varphi^{0}$ of the second class too,
we impose the gauge-fixing condition
\begin{equation}
\chi_{0} \definition h_{0} \approx 0 .
\end{equation}
For the total set of constraints
$\Phi^{M} \definition ( \varphi^{A}, \chi_{0} )$ with
$\varphi^{A} \definition ( \varphi^{0}, \varphi^{m}, \varphi_{1}^{0},
\varphi_{1}^{m}, \varphi_{2}^{0} )$, we have
\begin{equation}
{\rm Det}\left[ \Phi^{M}(\vecvar{x}), \Phi^{N}(\vecvar{x}')\right]
\propto 1 .
\end{equation}
The resulting functional integral expression is, therefore,
\begin{eqnarray}
Z & = &
\int{\cal D}\pi^{0}{\cal D}\pi^{m}{\cal D}\pi^{mn}
{\cal D}h_{0}{\cal D}h_{m}{\cal D}h_{mn}
\delta(\varphi^{A})\delta(\chi_{0}) \nonumber \\
& & \makebox[1cm]{}
\times\exp i\int d^{4}x \left[
\pi^{0}\dot{h}_{0} + \pi^{m}\dot{h}_{m} + \pi^{mn}\dot{h}_{mn} - H_{0}
\right] .
\end{eqnarray}

\section{Batalin-Fradkin extension}

In this section, following BF, we convert the second-class constraints
among (49) into first-class ones. This is done by introducing a
spin-1 auxiliary field (BF field) $\theta_{\mu}$ and its conjugate
momentum $\omega^{\mu}$ to modify the constraints $\varphi^{A}$ (49)
and the Hamiltonian $H$ (51).

The modification of the constraints is easily carried out as follows:
\begin{equation}
\begin{array}{rclcl}
\tilde{\varphi}^{0} & \definition & \varphi^{0} & \approx & 0 ,
\vspace{2mm} \\
\tilde{\varphi}^{m} & \definition & \varphi^{m} - \omega^{m}
& \approx & 0 , \vspace{2mm} \\
\tilde{\varphi}_{1}^{0} & \definition &
\varphi_{1}^{0} + \partial_{m}\omega^{m} + m^{2}\theta_{0}
& \approx & 0 , \vspace{2mm} \\
\tilde{\varphi}_{1}^{m} & \definition &
\varphi_{1}^{m} + m^{2}\theta^{m}
& \approx & 0 , \vspace{2mm} \\
\tilde{\varphi}_{2}^{0} & \definition &
\varphi_{2}^{0} - 2\partial^{m}\theta_{m} + 3\omega^{0}
& \approx & 0 .
\end{array}
\end{equation}
In fact all Poisson brackets between them vanish:
\begin{equation}
\left[ \tilde{\varphi}^{A}(\vecvar{x}),
\tilde{\varphi}^{B}(\vecvar{x}')\right] = 0 .
\end{equation}

The next step is to construct a modified Hamiltonian $\tilde{H}$
\begin{equation}
\tilde{H} = H + \Delta H(h,\pi;\theta,\omega)
\end{equation}
that has vanishing Poisson brackets with all the constraints (57) and
satisfies the boundary condition
\begin{equation}
\Delta H(h,\pi;0,0) = 0 .
\end{equation}
The conditions $\left[ \tilde{\varphi}^{A}, H\right] = 0$ give the
following equations:
\begin{eqnarray}
\lefteqn{
- \frac{\partial\Delta H}{\partial h_{0}} +
\partial^{m}\frac{\partial\Delta H}{\partial\pi^{m}}
= 0 ,
} \\
\lefteqn{
2\varphi_{1}^{m} -
\frac{\partial\Delta H}{\partial h_{m}} -
\partial^{m}\frac{\partial\Delta H}{\partial\pi^{0}} -
\partial^{m}\frac{\partial\Delta H}{\partial\pi^{kl}}\eta^{kl} +
\frac{\partial\Delta H}{\partial\theta_{m}} = 0 ,
} \\
\lefteqn{
- \partial_{m}\varphi_{1}^{m} +
\frac{m^{2}}{2}\varphi_{2}^{0} +
\left(\partial_{m}\partial^{m} - m^{2}\right)
\frac{\partial\Delta H}{\partial\pi^{kl}}\eta^{kl} -
\partial^{m}\partial^{n}\frac{\partial\Delta H}{\partial\pi^{mn}}
} \nonumber \\
& & \makebox[55mm]{} -
\partial_{m}\frac{\partial\Delta H}{\theta_{m}} +
m^{2}\frac{\partial\Delta H}{\partial\omega^{0}} = 0 , \\
\lefteqn{
\partial^{m}\varphi_{1}^{0} +
\frac{1}{2}\partial_{n}\partial^{n}\varphi^{m} +
\frac{1}{2}\partial^{m}\partial_{n}\varphi^{n} -
\partial_{n}\frac{\partial\Delta H}{h_{mn}} +
\partial^{m}\partial^{n}\frac{\partial\Delta H}{\pi^{n}}
} \nonumber \\
& & \makebox[55mm]{} +
m^{2}\frac{\partial\Delta H}{\pi_{m}} +
m^{2}\frac{\partial\Delta H}{\omega_{m}} = 0 , \\
\lefteqn{
\partial_{m}\varphi^{m} +
4\varphi_{1}^{0} -
\frac{\partial\Delta H}{h^{kl}}\eta^{kl} +
\partial^{m}\frac{\partial\Delta H}{\pi^{m}} -
2\partial^{m}\frac{\partial\Delta H}{\omega^{m}} -
3\frac{\partial\Delta H}{\theta_{0}} = 0 .
}
\end{eqnarray}
These equations can be solved successively:
Eqs. (62), (63), (64) and (65) determine in what form $\Delta H$ contains
the variables $\theta_{m}, \omega^{0}, \omega^{m}$ and $\theta_{0}$
respectively;
Eq. (61) just tells that the variable $h_{0}$ in $\Delta H$ is to be
combined with the variable $\pi^{m}$ to the form
$\pi^{m} - \partial^{m}h_{0}$.
After some lengthy but straightforward calculations we can find the
solution satisfying all the conditions (60) and ${\rm (61)}-{\rm (65)}$:
\begin{eqnarray}
\Delta H & = &
\varphi^{m}\left[
- \frac{1}{3}\left( 1 + \frac{2}{m^{2}}\partial_{n}\partial^{n}\right)
\partial_{m}\theta_{0} -
\frac{1}{2m^{2}}\partial^{n}\left(
\partial_{n}\omega_{m} + \partial_{m}\omega_{n}\right)\right]
\nonumber \\
& & +
\varphi_{1}^{0}\left[
\frac{1}{3}\left( 4 + \frac{2}{m^{2}}\partial_{m}\partial^{m}\right)
\theta_{0} +
\frac{1}{m^{2}}\partial_{m}\omega^{m}\right]
\nonumber \\
& & +
\varphi_{1}^{m}\left(
- 2\theta_{m} + \frac{1}{m^{2}}\partial_{m}\omega^{0}\right) +
\varphi_{2}^{0}\left( - \frac{1}{2}\omega^{0}\right)
\nonumber \\
& & -
\frac{1}{3}\partial_{m}\theta_{0}\partial^{m}\theta_{0} +
\frac{2}{3}m^{2}\theta_{0}^{\makebox[1mm]{}2} +
\theta_{0}\partial_{m}\omega^{m} -
\frac{1}{8m^{2}}\left(
\partial_{m}\omega_{n} - \partial_{n}\omega_{m}\right)^{2}
\nonumber \\
& & - m^{2}\theta_{m}\theta^{m} -
\frac{3}{4}\omega^{02} .
\end{eqnarray}

If we introduce new variables
\begin{equation}
\begin{array}{rcl}
\tilde{h}_{0} & \definition & h_{0} ,
\vspace{2mm} \\
\tilde{h}_{m} & \definition & \displaystyle
h_{m} + \theta_{m} - \frac{1}{m^{2}}\partial_{m}\omega^{0} ,
\vspace{2mm} \\
\tilde{h}_{mn} & \definition & \displaystyle
h_{mn} - \frac{1}{2m^{2}}\left(
\partial_{m}\omega_{n} + \partial_{n}\omega_{m}\right) -
\frac{1}{3}\left(
\eta_{mn} + \frac{2}{m^{2}}\partial_{m}\partial_{n}\right) \theta_{0} ,
\vspace{2mm} \\
\tilde{\pi}^{0} & \definition & \displaystyle
\pi^{0} - \partial^{m}\theta_{m} +
\frac{1}{m^{2}}\partial_{m}\partial^{m}\omega^{0} ,
\vspace{2mm} \\
\tilde{\pi}^{m} & \definition & \displaystyle
\pi^{m} - \left(
\eta^{mn} + \frac{1}{m^{2}}\partial^{m}\partial^{n}\right) \omega_{n} -
\frac{1}{3}\left(
3 + \frac{2}{m^{2}}\partial_{n}\partial^{n}\right) \partial^{m}\theta_{0} ,
\vspace{2mm} \\
\tilde{\pi}^{mn} & \definition & \displaystyle
\pi^{mn} - \eta^{mn}\partial^{l}\theta_{l} +
\left( \eta^{mn} + \frac{1}{m^{2}}\partial^{m}\partial^{n}\right)
\omega^{0} ,
\end{array}
\end{equation}
the modified constraints $\tilde{\varphi}^{A}$ and Hamiltonian
$\tilde{H}$ have more transparent expressions.
These are what we have through the simple replacement of the variables
$(h, \pi)$ by $(\tilde{h}, \tilde{\pi})$ in Eqs. (49) and (51):
\begin{eqnarray}
\tilde{\varphi}^{A} & = &
\varphi^{A} \left[ h, \pi \rightarrow \tilde{h}, \tilde{\pi}\right]
\approx 0 , \\
\tilde{H} & = & H \left[ h, \pi \rightarrow \tilde{h}, \tilde{\pi}\right] .
\end{eqnarray}
We thus constructed a first-class constrained system for a massive tensor
field.

\section{Gauge-fixings}

\subsection{{\it Unitary gauge}}

To eliminate the BF field, we impose the following gauge-fixing conditions
$\chi_{A} \definition
\left(\chi_{0}, \chi_{m}, \chi_{10}, \chi_{1m}, \chi_{20}\right)$:
\begin{equation}
\begin{array}{rclcl}
\chi_{0} & \definition & h_{0} & \approx & 0 , \\
\chi_{m} & \definition & \theta_{m} & \approx & 0 , \\
\chi_{10} & \definition & \omega^{0} & \approx & 0 , \\
\chi_{1m} & \definition & \omega^{m} & \approx & 0 , \\
\chi_{20} & \definition & \theta_{0} & \approx & 0 .
\end{array}
\end{equation}
Then the path integral
\begin{eqnarray}
Z & = &
\int{\cal D}\pi^{0}{\cal D}\pi^{m}{\cal D}\pi^{mn}{\cal D}h_{0}
{\cal D}h_{m}{\cal D}h_{mn}{\cal D}\omega^{0}{\cal D}\omega^{m}
{\cal D}\theta_{0}{\cal D}\theta_{m}
\delta(\tilde{\varphi}^{A})\delta(\chi_{A}) \nonumber \\
& & \times\exp i\int d^{4}x\left[
\pi^{0}\dot{h}_{0} + \pi^{m}\dot{h}_{m} + \pi^{mn}\dot{h}_{mn} +
\omega^{0}\dot{\theta}_{0} + \omega^{m}\dot{\theta}_{m} - \tilde{H}
\right]
\end{eqnarray}
immediately reduces to (56).
This shows that for a massive tensor field the original system (48)
is a gauge-fixed version of the extended system (68) and (69).

\subsection{{\it Massless-regular gauge}}

As the second gauge-fixing we impose
\begin{equation}
\begin{array}{rclcl}
\chi_{0} & \definition & h_{0} & \approx & 0 , \\
\chi_{m} & \definition & h_{m} & \approx & 0 , \\
\chi_{10} & \definition & \pi^{m}_{\makebox[1mm]{}m} + 3\omega^{0}
& \approx & 0 , \\
\chi_{1m} & \definition & \displaystyle
\partial^{n}h_{mn} - \frac{1}{2}\partial_{m}h^{n}_{\makebox[1mm]{}n}
& \approx & 0 , \\
\chi_{20} & \definition & \theta_{0} & \approx & 0 .
\end{array}
\end{equation}
In this case the path integral is given by
\begin{eqnarray}
Z & = &
\int{\cal D}\pi^{0}{\cal D}\pi^{m}{\cal D}\pi^{mn}{\cal D}h_{0}
{\cal D}h_{m}{\cal D}h_{mn}{\cal D}\omega^{0}{\cal D}\omega^{m}
{\cal D}\theta_{0}{\cal D}\theta_{m}
\delta(\tilde{\varphi}^{A})\delta(\chi_{A})\prod_{t}{\rm Det}M
\nonumber \\
& & \times
\exp i\int d^{4}x\left[
\pi^{0}\dot{h}_{0} + \pi^{m}\dot{h}_{m} + \pi^{mn}\dot{h}_{mn} +
\omega^{0}\dot{\theta}_{0} + \omega^{m}\dot{\theta}_{m} - \tilde{H}
\right] ,
\end{eqnarray}
where $M$ is what is defined by Eq. (38).
Carry out the integrations over
$h_{0}, h_{m}, \pi^{0}, \pi^{m}$ and $\theta_{0}$.
Do the following exponentiations
\begin{eqnarray}
\lefteqn{
\delta\left(
\partial_{m}\partial^{m}h^{n}_{\makebox[1mm]{}n} -
\partial^{m}\partial^{n}h_{mn} -
m^{2}h^{m}_{\makebox[1mm]{}m} +
\partial_{m}\omega^{m}\right)
} \nonumber \\
& & = \int{\cal D}h_{0}\exp i\int d^{4}x\left(
- \partial_{m}h_{0}\partial^{m}h^{n}_{\makebox[1mm]{}n} +
\partial^{m}h_{0}\partial^{n}h_{mn} -
m^{2}h_{0}h^{m}_{\makebox[1mm]{}m} +
h_{0}\partial_{m}\omega^{m}\right) ,
\nonumber \\
& & \\
\lefteqn{
\delta\left(
\partial_{n}\pi^{mn} + m^{2}\theta^{m}\right) =
\int{\cal D}h_{m}\exp i\int d^{4}x\left[
- \pi^{mn}\left(
\partial_{m}h_{n} + \partial_{n}h_{m}\right) +
2m^{2}h_{m}\theta^{m}\right] ,
}
\nonumber \\
& & \\
\lefteqn{
\delta\left( -2\partial^{m}\theta_{m}\right) =
\int{\cal D}\lambda_{0}\exp i\int d^{4}x
\left( 2\partial_{m}\lambda_{0}\theta^{m}\right) .
}
\end{eqnarray}
And then integrate over $\omega^{0}, \pi^{mn}$ and $\theta_{m}$.
Write $\theta_{0}$ and $\theta_{m}$ over $- \frac{1}{m^{2}}\lambda_{0}$
and $\frac{1}{2m^{2}}\omega_{m}$ respectively.
The final result we arrive at is
\begin{eqnarray}
Z & = &
\int{\cal D}h_{0}{\cal D}h_{m}{\cal D}h_{mn}
{\cal D}\theta_{0}{\cal D}\theta_{m}
\delta\left(
\partial^{m}h_{m} - \frac{1}{2}\dot{h}^{m}_{\makebox[1mm]{}m}
\right)
\delta\left(
\partial^{n}h_{mn} - \frac{1}{2}\partial_{m}h^{n}_{\makebox[1mm]{}n}
\right)
\prod_{t}{\rm Det}M \nonumber \\
& & \makebox[2cm]{}
\times\exp i\int d^{4}x\left[
L[ h, m=0 ] + R \right] ,
\end{eqnarray}
where
\begin{eqnarray}
R & \definition &
- \frac{m^{2}}{2}\left[\left(
h_{mn} - \partial_{m}\theta_{n} - \partial_{n}\theta_{m}\right)^{2} -
\left( h^{m}_{\makebox[1mm]{}m} - 2\partial^{m}\theta_{m}\right)^{2}
\right.\nonumber \\
& & \left.\makebox[2cm]{} +
2h_{0}\left( h^{m}_{\makebox[1mm]{}m} - 2\partial^{m}\theta_{m}\right) -
2\left( h_{m} - \dot{\theta}_{m} - \partial_{m}\theta_{0}\right)^{2}
\right] .
\end{eqnarray}
When the mass $m$ goes to zero, the fields $\theta_{0}$ and $\theta_{m}$
become redundant and the path integral (77) tends to the expression
obtained by setting $f_{\mu} = 0$ in Eq, (43).
It has turned out that the BF-extended theory of massive tensor field
equipped with the gauge-fixing conditions (72) does smoothly reduce to
the massless tensor theory of Coulomb-like gauge.

The expression (77) is not covariant.
For the purpose of having a covariant expression,
it may be useful to rewrite Eq. (77) as
\begin{eqnarray}
Z & = &
\int{\cal D}h_{\mu\nu}{\cal D}\theta_{\mu}
\underline{
\delta\left(
\partial^{n}h_{0n} - \frac{1}{2}\partial_{0}h^{n}_{\makebox[1mm]{}n}
\right)
\delta\left(
\partial^{n}h_{mn} - \frac{1}{2}\partial_{m}h^{n}_{\makebox[1mm]{}n}
\right)
\prod_{t}{\rm Det}M
}
\nonumber \\
& & \times\exp i \int d^{4}x\left[
L[ h, m=0 ] -
\frac{m^{2}}{2}\left(\left(
h_{\mu\nu} - \partial_{\mu}\theta_{\nu} - \partial_{\nu}\theta_{\mu}
\right)^{2} -
\left(
h^{\mu}_{\makebox[1mm]{}\mu} - 2\partial^{\mu}\theta_{\mu}
\right)^{2}\right)
\right.\nonumber \\
& & \left.\makebox[3cm]{}
\begin{array}{c}{}\\{}\end{array}
\underline{
- 2m^{2}\partial_{0}\theta_{0}\left(
h^{m}_{\makebox[1mm]{}m} - 2\partial^{m}\theta_{m}\right)
}
\right] .
\end{eqnarray}
In this expression the non-covariant factors are underlined,
all of which turn out to be related to gauge-fixings.
To find a covariant expression by setting some suitable gauge conditions
is left to be solved.

\section{Summary and discussion}

We have applied BF algorithm to a massive tensor field theory.
By introducing an auxiliary vector field we have been able to convert the
original second-class constrained system into a first-class one.
Two gauge-fixings have been discussed:
one is what recovers the original massive theory;
the other is the one that has a smooth massless limit.

Our final expressions are still non-covariant.
To find some suitable gauge-fixing conditions that lead to covariant
expressions remains to be solved.
Furthermore, our formalism has been restricted to the linear free theory.
To construct a complete nonlinear theory such as smoothly reduces to
general relativity in the massless limit is another big problem.

Investigation in this direction seems to be useful for solving the
infrared problem in quantum gravity.

\section*{Acknowledgments}

The author would like to thank Minoru Hirayama for discussions.

\end{document}